# APPLICATION OF CORONA DISCHARGE FOR POLING FERROELECTRIC AND NONLINEAR OPTICAL POLYMERS

S. N. Fedosov, A .E. Sergeeva, T. A. Revenyuk, A. F. Butenko
*Odessa National Academy of Food Technologies, Odessa, Ukraine*

## *I. Introduction*

Polar species in ferroelectric and nonlinear optical polymers must be oriented in one direction by application of an external electric field in order to ensure required working characteristics of the materials. The corresponding process called poling or charging is an important step in obtaining desired properties. The most advanced is the corona charging method applied earlier in electrophotography [1], in electrets [2] and in electrostatic filters [3]. Due to its versatility, the corona method allows to optimize poling by proper selection of the poling mode, polarity and temperature.

A corona is a self-sustainable discharge occurring if a sufficiently high voltage is applied to asymmetric electrodes such as a point and a plate [4]. In the ionization area near the point, depending on corona polarity, either positive or negative ions are produced. The ions bombard a non-metallized sample surface, so the electrode is not required on this side. The process takes place at atmospheric pressure, so the ions have an average thermal energy. The ions do not penetrate into the bulk, but transfer their charge and leave the surface as neutral atoms or molecules. The excess charge either resides on the surface, or injected in the bulk. The advantages of the corona charging are as follows: (a) poling can be performed without deposited electrodes; (b) the higher fields can be achieved than in the case of sandwich poling, and (c) thin films can be poled in spite of defects, because breakdowns are limited only to small sample areas.

In practice of the charging, the simple "point-to-plane" geometry has been gradually replaced by a corona triode with a metal grid introduced between the point and the sample. Apart from charging, the corona triode was used to study dynamics of poling and charge transport phenomenon in polymers [5-10]. This paper gives a short review of corona poling applied for charging ferroelectrics and nonlinear optical polar polymers with the emphasis on using the constant current modification of corona poling.

## *II. Experimental setups*

In this paper we consider only corona triode methods, because the "point-to-plane" technique does not provide the required quality of poling. There are four main modifications of the corona triode (Fig. 1). In the simplest mode I, corona and grid voltages are controlled independently by power supplies 5 and 6 and kept constant. The elements 4, 8, 9, 10 shown in Fig. 1 are not used in the mode I. One can measure the DC poling current (7), but cannot separate its components. The highest surface potential is approximately equal to the grid voltage. If either the sample, or the grid is made vibrating (mode II, element 4 is added), one can observe the dynamics of the surface potential by the modified Kelvin method of measuring the AC current component (8). In the mode III (constant current corona poling or CCCP mode) additionally to the vibrating grid the feedback 10 is introduced to control the corona voltage 6 in order to keep the poling current constant. So, all poling parameters can be measured and controlled [11-16]. In the latest version of the CCCP triode (mode IV) [17] not the corona, but the grid voltage is adjusted continuously through the feedback 9 for keeping the current constant. The sample potential can be found without using the vibrating capacitor, so elements 4 and 10 are excluded. The surface potential is found from the graduation curve obtained beforehand by substitution a real sample with a metal disk of the same area. Detailed description of the method is given in review papers [6, 7].

To fix corona and grid voltages, as well as point-to-grid and grid-to-sample distances we recommend to measure current-voltage characteristics under different conditions (Fig. 2). Since the corona appears at about 5 kV, the grid potential should not exceed this value, otherwise the grid itself produces a parasitic corona. At the same time, the difference between the corona voltage and the grid voltage must not be smaller than 5 kV, or it should be kept constant to ensure a stable discharge at different grid voltages. In most cases the value up to 4 kV and 12-16 kV are suitable as the grid and the corona voltages. After completion of poling, a short-circuiting of the sample can be performed by grounding the grid and changing corona polarity to the opposite one. This step is desirable in many cases in order to remove excess surface charge.

### *III. Corona poling of NLO polymers*

Polymers for nonlinear optics gained interest because of the strong nonlinearity of molecular chromophores as guests, side groups, or main-chain segments. In order to obtain a non-centrosymmetric material, it is necessary to perform a poling process. The suitable method for NLO polymers is corona poling [18-20] applied near the glass transition temperature $T_g$ to allow orientation of dye molecules in the field. Cooling the sample with the corona still applied freezes the orientation of dipoles. Unfortunately, the corona charging experience was not used in full in the NLO field where most research was carried out with a poorly controlled point-to-plane mode of the method. For example, the surface potential after poling was only 50 V, while the corona electrode was kept at 5 kV [19]. We know from our experience that one should use voltages of 8-20 kV to obtain the stable corona [10, 20].

Only recently the corona triode was used to pole NLO polymers [21]. Poling time is an important parameter in this case. The time constant for alignment of dye molecules at $T_g$ is of the order of milliseconds [22]. However, the SHG signal continued to increase even after one hour of poling. This indicates that the surface potential increased slowly during all this time.

### *IV. Corona poling of ferroelectric polymers*

Due to high piezo- and pyroelectricity, ferroelectric polymers may replace conventional inorganic materials in sensors and transducers. Therefore, such materials as PVDF, P(VDF-TrFE), P(VDF-TFE) have attracted attention recently. To induce orientation of dipoles, the polymers are poled usually in a corona [6-17], because high electric fields could be achieved without breakdown in this case. Using a constant current corona triode we have found that the initial poling and switching consisted of three stages, each one corresponded to a definite part of the potential - time curve [7, 8, 14-16, 23, 24]. The fast increase in surface potential was observed at the first stage indicating that the capacitive component prevailed in the poling current. At the second stage there was a plateau at the voltage-time curve related, most probably, to switching of the ferroelectric component of polarization. The surface potential again increased sharply at the third stage when switching is completed. The plateau was not seen if poling is repeated.

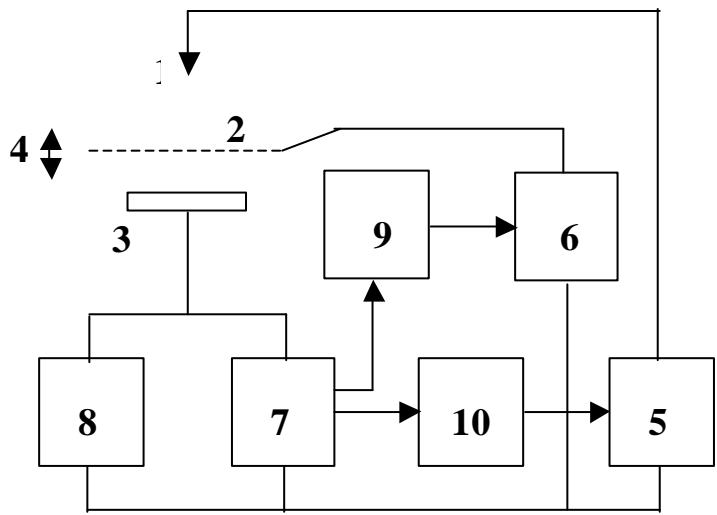

**Figure 1.** Block-diagram of the corona setup. **1** – the corona electrode, **2** – the grid, **3** – the sample, **4** – vibration of the grid, **5** – high voltage DC power supply for corona, **6** – power supply for the grid, **7** – DC component of the poling current, **8** – AC component of the poling current, **9** – the feedback circuit to control the grid power supply, **10** – the feedback circuit to control the corona power supply.

Polarization $P$ in ferroelectric polymers depends nonlinearly on the field $E$, so the $P(E)$ function is presented by a hysteresis loop. From this curve, the remanent polarization $P_r$ and the coercive field $E_c$ can be found. We studied polarization and hysteresis phenomena in biaxially and uniaxially stretched PVDF films. The $P(E)$ dependence was obtained from the kinetic of the surface potential during CCCP [8]. For the biaxially stretched PVDF we found $E_c$=100 MV/m and $P_r$ = 64 mC/m$^2$. We separated electronic, dipolar and ferroelectric components of the dielectric constant and obtained $e_e = 2$, $e_d = 7$, $e_f = 95$. In case of uniaxially stretched films we have found $e_f = 40$, $P_r = 42$ mC/m$^2$ and $E_c$ =48 MV/m [25,26], all values lower than those for biaxially stretched PVDF.

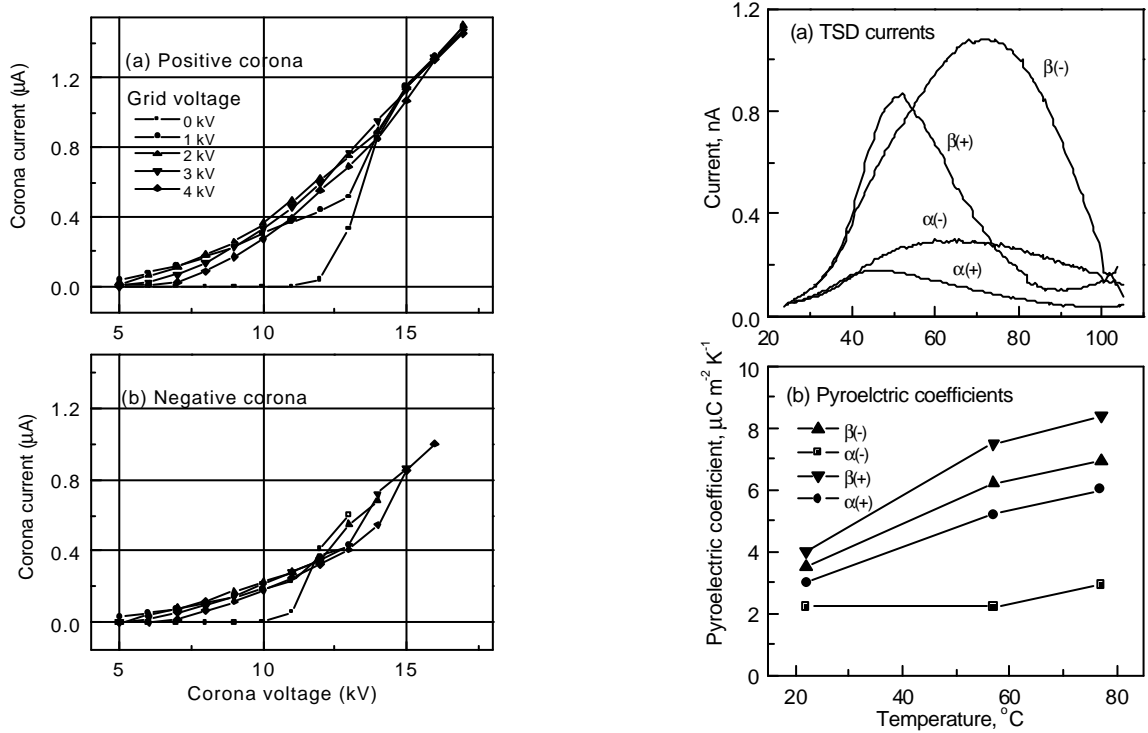

**Figure 2.** Current-voltage characteristics of positive (a) and negative (b) coronas.

**Figure 3.** TSD currents (a) of negatively and positively corona charged alpha- and beta-PVDF and corresponding pyroelectric coefficients measured by the dynamic method.

We studied polarization uniformity in corona poled samples of P(VDF-TFE) 95:5 copolymer (Plastpolymer, Russia) containing 90% of ferroelectric beta phase [27]. Polarization profiles were measured by the piezoelectrically induced pressure step (PPS) method [28]. It has been found that the residual polarization is distributed nonuniformly in samples poled by CCCP method independently on poling temperature. Nonuniformity of polarization was caused by nonuniform distribution of the poling field that, in its turn, was attributed to injection of negative charge during poling [29].

Efficiency of corona poling depends on corona polarity. We compared TSD currents and pyroelectric coefficients of PVDF samples containing preferentially either polar beta phase or non-polar alpha phase and poled in either positive or negative corona. Corresponding results are presented in Fig. 3. It is seen that beta samples are poled to the higher values and show higher pyroelectricity that the alpha samples. Judging by values of the pyrocoefficients, poling in a positive corona is more efficient than in a negative one, probably because the positive charges are not easily injected into the bulk, as do the negative ones.

*V. Final remarks*

Corona poling is a powerful method to produce residual polarization in polar polymers. The advantages of corona charging are that samples can be poled without deposited electrodes, higher fields can be achieved in corona than in case of sandwich contact poling, thin films can be poled in spite of defects. Moreover, the valuable information about charge transport, storage and polarization can be obtained during poling in a corona triode. The constant current method facilitates analysis of experimental results and gives additional information on injection and drift of carriers, the hysteresis phenomena and polarization buildup.